\documentclass[12pt]{article}
\def\be {\begin{equation}}
\def\ee {\end{equation}}
\def\ba {\begin{eqnarray}}
\def\ea {\end{eqnarray}}

\def\bi {\begin{itemize}}
\def\ei {\end{itemize}}
\begin{document}
\def\bea{\begin{eqnarray}}
\def\eea{\end{eqnarray}}
\title{Spacing of the entropy spectrum for KS Black hole in Ho$\check{\textbf{r}}$ava-Lifshitz
 gravity}
\author{  \textbf{M. R. Setare} \thanks{%
E-mail: rezakord@ipm.ir} \\ \textbf{ D. Momeni} \thanks{%
E-mail:dmomeni@phymail.ut.ac.ir
 }\\{Department of Science, Payame Noor University, Bijar, Iran} \\
}

\maketitle

\begin{abstract}
In this paper we present the spectrum of entropy/area for
Kehagias-Sfetsos (KS) black hole in
Ho$\check{\textbf{r}}$ava-Lifshitz (HL)gravity via quasi-normal
modes (QNM) approach. We show that in the massive case the mass
parameter $\mu$ disappears in the entropy spectrum and only the
quasinormal  modes modified by a term which is proportional to the
mass square term. Our calculations show that the charge like
parameter $\frac{1}{2\omega}=Q^{2}$ appears in the surface gravity
and our calculations can be applied to any spherically symmetric
spacetime which has only one physically acceptable horizon. Our
main difference
 between our calculations and what was done in \cite{1} is that the
portion of charge and mass is included explicitly in the surface
gravity and consequently in the QNM expression. Since the
imaginary part of the QNM is related to the
 adiabatic invariance of the system and also to the entropy,
surprisingly the mass parameter do not appear in the entropy
spectrum. Our conclusion supported by some acclaims about that the
scalar field parameters (charges ) can not change the fundamental
parameters in the 4-dimensional black holes.
\end{abstract}

\section{I:\label{sec:level1}Introduction}
The quantization of the area of a black hole in general relativity
(GR) has an old history. This topic return to the Bekenstein works
on the BH physics \cite{2'}. Without no doubt the first acclaim
about the existence of an upper bound for the BH entropy and the
corresponding analogous as a holographic model belongs to the
Bekenstein. The Hawking radiation \cite{3'} was discussed in many
papers and by different authors both in the context of the GR and
the alternative gravity models.\\
Recently Ho$\check{\textbf{r}}$ava presents a new non
relativistic model for Quantum gravity which has renormalizable
and has a good UV limit for propagators \cite{4',5',6'}. There is
no unique vacuum solution for HL theory. Also latterly, some
cylindrical and plane solutions was obtained \cite{5,6}. More
recently Majhi \cite{1} discussed Hawking radiation and spectrum
of entropy/area for the Cai-Cao and Ohta \cite{2} spherically
symmetric static black hole solution in
Ho$\check{\textbf{r}}$ava-Lifshitz theory by  using Tunneling
formalism \cite{3} and QNM \cite{4}.  According to the Blas et al
arguments \cite{7}, it seems that this model must be modified by
some terms to avoiding from strong coupling, instabilities,
dynamical in consistencies and unphysical extra mode. One of the
first exact solutions for this modified version is the work of
Kiritsis\cite{8}. Indeed the Kiritsis work contains some previous
families of exact solutions as a special sub class and has a good
asymptotic behaviors. The explicit form of exact solution for
this modified version deal with some algebraic quadratures and
lead finally to an implicit static spherically symmetric metric.
But no doubt this solution generic, avoids from the trouble
problems which occur in the original version of HL. The
thermodynamics of the HL black holes was discussed by A. Wang et.
al \cite{11} and specifically for KS solution by M. Wang, et. al
\cite{12}. Later the Area spectrum for different BHs
calculated \cite{9}.\\
The quantization of the black hole area in the framework of
Einstein gravity, has been considered \cite{9', 10'}, as a result
of the absorption of a quasi-normal mode excitation. The
quasi-normal modes of black holes are the characteristic, ringing
frequencies which result from their perturbations \cite{11'} and
provide a unique signature of these objects \cite{12'}, possible
to be observed in gravitational waves. In this short paper, we
focused on the KS solution \cite{10} and following the QNM method
we derived an expression for entropy/area spectrum.
\section{II:KS black hole solution in HL theory}
Following from the ADM decomposition of the metric \cite{22}, and
the Einstein equations, the fundamental objects of interest are
the fields $N(t,x),N_{i}(t,x),g_{ij}(t,x)$ corresponding to the
\emph{lapse }, \emph{shift} and \emph{spatial metric} of the ADM
decomposition.
 In the $(3 + 1)$-dimensional ADM formalism, where the
metric can be written as
 \begin{eqnarray}\nonumber
ds^2=-N^2dt^2+g_{ij}(dx^{i}+N^{i}dt)(dx^{j}+N^{j}dt)
\end{eqnarray}
and for a spacelike hypersurface with a fixed time, its extrinsic
curvature $K_{ij}$ is

\begin{eqnarray}\nonumber
K_{ij}=\frac{1}{2N}(\dot{g_{ij}}-\nabla_{i}N_{j}-\nabla_{j}N_{i})
\end{eqnarray}
where a dot denotes a derivative with respect to t and covariant
derivatives defined with respect to the spatial metric $g_{ij}$,
the action of Ho$\check{\textbf{r}}$ava-Lifshitz theory  for $z=3$
is
\begin{eqnarray}\nonumber
S=\int_{M} dtd^{3}x\sqrt{g} N(\mathcal{L}_{K} - \mathcal{L}_{V} )
\end{eqnarray}
we define the space-covariant derivative on a covector $v_{i}$ as
$\nabla_{i}v_{j}\equiv \partial_{i}v_{j}-\Gamma_{ij}^{l}v_{l}$
where $\Gamma_{ij}^{l}$ is the spatial Christoffel symbol. $g$ is
the determinant of the 3-metric and $N = N(t)$ is a dimensionless
homogeneous gauge field. The kinetic term is
\begin{eqnarray}\nonumber
\mathcal{L}_{K}=\frac{2}{\kappa^2}\mathcal{O}_{K}=\frac{2}{\kappa^2}(K_{ij}K^{ij}-\lambda
K^2)
\end{eqnarray}
Here $N_{i} $ is a gauge field with scaling dimension $[N_{i}] =
z -
1$.\\
The \emph{'potential'} term $\mathcal{L}_{V}$ of the
$(3+1)$-dimensional theory is determined by the \emph{principle of
detailed balance }\cite{4'}, requiring $\mathcal{L}_{V}$ to
follow, in a precise way, from the gradient flow generated by a
3-dimensional action $W_{g}$. This principle was applied to
gravity with the result that the number of possible terms in
$\mathcal{L}_{V} $ are drastically reduced with respect to the
broad choice available in an '\emph{potential} is
\begin{eqnarray}\nonumber
\mathcal{L}_{V}=\alpha_{6}C_{ij}C^{ij} -
\alpha_{5}\epsilon_{l}^{ij} R_{im}\nabla_{j}R^{ml} + \alpha_{4}
[R_{ij}R^{ij}- \frac{4\lambda-1}{4(3\lambda-1)} R^2]
+\alpha_{2}(R - 3\Lambda_{W})
\end{eqnarray}
Where in it $C_{ij}$ is the \emph{Cotton }tensor \cite{4'} which
is defined as,
\begin{eqnarray}\nonumber
C^{ij}=\epsilon^{kl(i}\nabla_{k}R^{j)}_{l}
\end{eqnarray}
The kinetic term could be rewritten in terms of the \emph{de Witt
metric} as:
\begin{eqnarray}\nonumber
\mathcal{L}_{K}=\frac{2}{\kappa^2}K_{ij}G^{ijkl}K_{kl}
\end{eqnarray}
Where we have introduced the \emph{de Witt metric}
\begin{eqnarray}\nonumber
G^{ijkl}=\frac{1}{2}(g^{ik}g^{jl}+g^{il}g^{jk})-\lambda
g^{ij}g^{kl}
\end{eqnarray}
The inverse of this metric is given by
\begin{eqnarray}\nonumber
G_{ijkl}=\frac{1}{2}(g_{ik}g_{jl}+g_{il}g_{jk})-\tilde{\lambda}g_{ij}g_{kl}\\\nonumber
\tilde{\lambda}=\frac{\lambda}{3\lambda-1}
\end{eqnarray}
Inspired by methods used in quantum critical systems and non
equilibrium critical phenomena, Ho$\check{\textbf{r}}$ava
restricts the large class of possible potentials using the
principle of detailed balance outlined above. This requires that
the potential term takes the form
\begin{eqnarray}\nonumber
\mathcal{L}_{V}=\frac{\kappa^2}{8}E^{ij}G_{ijkl}E^{kl}
\end{eqnarray}
Note that by constructing $E^{ij} $ as a functional derivative it
automatically transverse within the foliation slice,
$\nabla_{i}E^{ij}=0$. The equations of motion were  obtained in
\cite{23}. KS BH is a static spherically symmetric solution for HL
theory which contains 2 parameter , one mass like parameter  $m$
and a parameter which controls the escape from a naked singularity
$\omega$ and satisfies \cite{10}
\begin{eqnarray}\nonumber
\omega m^{2}\geq\frac{1}{2}
\end{eqnarray}
In the usual spherical coordinates  $(t,r,\theta,\phi)$ and in the
Schwarzschild's gauge the metric reads:
\begin{eqnarray}
ds^{2}=diag(-f,\frac{1}{f},r^{2}\Sigma_{2})
\end{eqnarray}
where in it the metric gauge function is
\begin{eqnarray}
f=1+\omega r^{2}-\sqrt{\omega^2 r^{4}+4 m \omega r}
\end{eqnarray}
and is  $\Sigma_{2}$ the surface element on a unit 2- sphere. As
motivated by Sekiwa "\emph{it is obvious that $1/2\omega$ is
equivalent to $Q^{2}$ and this means that we could view
$1/2\omega$ as a charge in some degree}"\cite{12,13}. Thus The
outer and inner event horizon can be compared with the outer and
inner event horizon of Reissner-Nordstrom black hole. Essentially
as claimed  by the founders of the KS, this solution
"\emph{represents the analog of the Schwarzschild solution of
GR}".

 \section{III:QNM formalism for  scalar field in spherically
 symmetric KS backgrounds}

QNM concept is based on Bohr's correspondence principle (1923):
\emph{transition frequencies at large quantum numbers should
equal classical oscillation frequencies}. Hence, we are
interested in the asymptotic behavior (i.e., the $n\rightarrow
\infty$ limit) of the ringing frequencies. These are \emph{the
highly damped black-hole oscillations frequencies, which are
compatible with the statement quantum transitions do not take
}(See, for example, \cite{15}). We consider a massive scalar field
with a typical mass $\mu$ satisfying the wave equation
$(\nabla_{\alpha}\nabla^{\alpha}+\mu^{2})\phi=0$. The scalar
perturbation fields for a massive particle outside the
spherically symmetric black hole (1) are governed by a
one-dimensional Schrudinger-like wave equation\cite{4}:
\footnote{Assuming a time dependence of the form $e^{i\omega
t}$and we decompose the scalar field as
$\phi=\frac{\Psi}{r}Y_{lm}(\theta,\varphi)e^{i\omega t}$}

\begin{eqnarray}
-\frac{d^{2}\Psi}{dr^{2}_{*}}+V(r)\Psi=\omega^{2}\Psi
\end{eqnarray}
where the tortoise radial coordinate $r_{*}$ is related to the
spatial radius $r$ (using the metric function (2)) by
\begin{eqnarray}\nonumber
r_{*}=\int\frac{dr}{f}
\end{eqnarray}
and the Potential term is given by
\begin{eqnarray}
V(r)=f(\frac{l(l+1)}{r^{2}}+\frac{f'}{r}-\mu^{2})
\end{eqnarray}
$l$ is the multipole moment index.
 For electromagnetic, and
gravitational perturbations see \cite{4}. For solving the ODE (3)
we must adopted a suitable boundary conditions. As one can observe
that, this equation cannot be solved exactly and in general only
we can determine aseries solution or a \emph{Poincare}'s
asymptotic solution which is very good for our QNM purposes. One
method is due to Medved, et. al \cite{4}. Since the potential
function (4) vanishes at the horizon located at the positive and
real root of the algebraic equation $f=0$ or the limiting point
of the tortoise coordinate $r_{*}\rightarrow-\infty$, and at
spatial infinity $r\rightarrow\infty $ or equivalently at
$r_{*}\rightarrow\infty $, it is so adequate to define the QNMs to
be those  modes for which we have purely ingoing plane wave at
the horizon and no wave at spatial infinity. The mathematical
expression which addressed correctly to this situation is,
\begin{eqnarray}
\Psi_{QNM}= e^{i\omega r_{*}} \hspace{0.5cm} at \hspace{0.5cm} r_{*}\rightarrow -\infty\\
\Psi_{QNM}= 0 \hspace{0.5cm} at \hspace{0.5cm} r_{*}\rightarrow
-\infty
\end{eqnarray}
The remaining part of the calculations is straightforward. We
solve the ODE (3) near horizon $r=h$ and comparing the solution
at the asymptotic region with (5),(6). By definition of the new
radial coordinate $z=r-h$ and represnting the exact solution for
the (3) in terms of  the value of the surface gravity as
$\kappa=\frac{1}{2}f'(h)$. We consider first massless limit of the
potential barrier (4).

\subsection {Massless scalar fields $\mu^{2}=0$}
If we set the mass parameter $\mu=0$ in potential function (4), we
can write the following expression for (3).\footnote{U is the
confluent hypergeometric function has the integral representation
\begin{eqnarray}\nonumber
U[a,b,z]=\frac{1}{\Gamma(a)}\int_{0}^{\infty}e^{-zt}t^{a-1}(1+t)^{b-a-1}dt
\end{eqnarray}}The degenerate hypergeometric functions $\Phi(a,b;x)$ and
$\psi(a,b;x)$ are solutions of the degenerate hypergeometric
equation. In the case $b$ is not negative integer, the function
$\Phi(a,b;x)$  can be represented as Kummer's series.

\begin{eqnarray}
\Psi=z^{i\frac{\omega}{2\kappa}}U[\alpha,\beta,\gamma
z]\\\nonumber \kappa=2\omega\frac{h-2m}{1+\omega h^{2}}\\\nonumber
h=m+\sqrt{m^{2}-Q^{2}}\\\nonumber
\alpha=\frac{1}{4}(2-i/\kappa(\delta(2\kappa
h+l(l+1))-2\omega))\\\nonumber
\beta=1+i\frac{\omega}{\kappa}\\\nonumber\gamma=2i\sqrt{h+l(l+1)/\kappa}/h^{3/2}\\\nonumber
\delta=\sqrt{\frac{\kappa}{h(h\kappa+l(l+1))}}
\end{eqnarray}
We note here that for the asymptotic boundary conditions the
second function $LaguerreL[n, a, x]$ gives the generalized
Laguerre polynomial and only if $a=0$ ,$LaguerreL[n, x]$ is an
entire function of $x$ with no branch cut discontinuities. Here
is the series expansion around $z=0$ to order 1 for generalized
Laguerre polynomials $LaguerreL[n, a, bz]$.
\begin{eqnarray}\nonumber
LaguerreL[n, a, 0]-b LaguerreL[-1 + n, 1 + a, 0] z +Q[z^{2}]
\end{eqnarray}

Since in our solution $a$ does not vanish at all, thus the only
way to avoiding from a un desirable branch cut discontinuities is
that setting the $c_{2}=0$. It seems that the author of the
\cite{1} completely ignore from this function. But we show that
this term must be ignored by consideration of the un-physical
discontinuity of the wave solution near the horizon. We know that
the value of the field function must be remained finite on the
horizon. The only quantity which may be diverge on this surface
is the value of the stress-energy tensor. In the  limit
$z\rightarrow 0$ using from the Asymptotic expansion series
expression for the Hypergeometric functions
\begin{eqnarray}\nonumber
\Phi(a,b;x)=\frac{\Gamma(b)}{\Gamma(a)}e^{x}x^{a-b}[\sum^{N}_{n=0}\frac{(b-a)_{n}(1-a)_{n}}{n!}x^{-n}+\varepsilon],x>0\\\nonumber
\Phi(a,b;x)=\frac{\Gamma(b)}{\Gamma(b-a)}(-x)^{-a}[\sum^{N}_{n=0}\frac{(a)_{n}(a-b+1)_{n}}{n!}(-x)^{-n}+\varepsilon],x<0\\\nonumber
\varepsilon=O(x^{-N-1})
 \end{eqnarray}

and comparing the solution with our desired boundary conditions
(5),(6)
\begin{eqnarray}
\Psi=c'_{1}e^{-1/2(\beta-1)}
\frac{\Gamma(\beta-1)}{\alpha(\omega)}+c'_{2}e^{1/2(\beta-1)}
\frac{\Gamma(1-\beta)}{\alpha(-\omega)}
\end{eqnarray}
 we lead to the following relation
for the QNMs :
\begin{eqnarray}
\omega_{n}=\delta(\kappa h+l(l+1)/2)+i\kappa(2 n+1)
\end{eqnarray}
Immediately the imaginary part of the frequency of the QNMs is
\begin{eqnarray}
Im(\nu_{n})=(2n+1)\kappa=2\pi(2n+1)\frac{T_{H}}{\hbar},
\end{eqnarray}
where $T_{H}$ is the Hawking temperature of the black hole.
Finally by calculating the adiabatic invariant quantity and using
of the first law of thermodynamics, the Bohr-Sommerfield
quantization rule we can write the next formula for the spacing
of the entropy spectrum
\begin{eqnarray}
S_{n}=4\pi n
\end{eqnarray}
\subsection {Massive scalar fields $\mu^{2}\neq 0$}
In case of the massive particle, the only change that must be done
is a redefinition of the the set of the parameters
$\alpha,\beta,...$ in (7). By a simple computations similar to
which done for a massless one, we observe that the form of the
scalar perturbations (7) must be written as
\begin{eqnarray}
\Psi=c_{1}z^{i\frac{\omega}{2\kappa}}\Phi[\alpha',\beta,\gamma
z]++ c_{2}z^{i\frac{\omega}{2\kappa}}
LaguerreL[\alpha',\beta-1,\gamma z]
\end{eqnarray}
where in it the new shifted parameter $\alpha$ is defined by
\begin{eqnarray}
\alpha'=\frac{i\sqrt{\kappa}(h^{2}\mu^{2}-l(l+1)-2h\kappa)+2\sqrt{h(h\kappa+l(l+1))}(\kappa+i\omega)}{4\kappa\sqrt{h(h\kappa+l(l+1))}}\\\nonumber
\alpha'=\alpha+i\frac{\delta h^{2}\mu^2}{4\kappa}
\end{eqnarray}
With a same discussion as (7) we can set $c_{2}=0$, and another
parameters coincide with (7). Instead of the QNMs (9) we have
another  slightly different equation as:
\begin{eqnarray}
\omega_{n}=\delta(\kappa h+l(l+1)/2)-\frac{\delta
h^{2}\mu^2}{2}+i\kappa(2 n+1)
\end{eqnarray}
which obviously recovers the massless equation (9). Following the
next steps as was written for massless case show that there is no
portion of the mass parameter in the spectrum of the entropy. In
the other hand, from a pure quantum mechanical point of view, the
mass of the particle only changed the zero point energy and not
the QNMs.

We mention here that the equispaced property remains unchanged
even if we use from  the tunneling mechanism as was shown for
topological black holes in  HL theory \cite{1}. Although the exact
value of the spacing in these two different approaches   does not
coincide, but
their order of magnitudes are same.\\
\section{IV:Conclusion}
In the present short letter, we have studied the entropy spectrum
associated with the   black hole event horizon  for KS black hole
solutions in HL theory. We derived the spectrum of entropy/area of
the new kind of the black hole in context of the HL theory, KS
black hole via the QNM approach for massless and massive scalar
fields pertubations. Explicitly we showed that the entropy
spectrum was equispaced in the large quantum number limit as
usually happens for Einstein gravity and Einstein-Gauss- Bonnet
gravity. On the other hand, since the entropy was not
proportional to the area, the area spectrum was not equispaced.
Consequently, it has a similarity with the Einstein-Gauss-Bonnet
theory, rather than the usual Einstein gravity. Also we derived
the general formula for the QNM. This equation contains two
Quantum numbers: first the principal quantum number $n$ and second
the multipole quantum number $l$. The spectrum of the eigen
frequencies correspond to the QNM are very like to the well known
spectrum of a 3-dimensional spherically symmetric simple harmonic
oscilator. As we know that if a simple charged  harmonic
oscillator exerted to a uniformly electric field in a specified
direction the energy spectrum of the oscillator changes only up to
order of a term which is proportional to the square of the field
strength. This simple exact result coincides to the perturbative
result for any order of the correction. The simplest contribution
of the angular quantum numbers arises from the monopole term
$l=0$. In \cite{1} the results were presented only for this
special case and nothing was stated about the contribution of
another higher multipole terms. As we can see from  (9) and (14)
if the multipole $l$ increases, the $\delta$ factor decreases and
the variation of the QNM totally monotonically increases and we
can argue that any multipole term bigger than monopole ones,
increases the QNM. Thus we focused only on the monopole $l=0$
term, we leak more information of the classical spectrum of the
oscillations.

\section{Acknowledgement}
D.Momeni thanks  Anzhong Wang for illuminating discussions.


\begin{thebibliography}{35}
\bibitem{1}
B. R. Majhi,"Hawking radiation and black hole spectroscopy in
Ho$\check{\textbf{r}}$ava-Lifshitz
gravity,"[arXiv:0911.3239[hep-th]].
\bibitem{2'}J. D. Bekenstein, Phys. Rev. D 7, 8, (1973).
\bibitem{3'}S. W. Hawking, "Particle Creation By Black Holes," Commun. Math.
Phys. 43, 199 (1975) [Erratum-ibid. 46, 206 (1976)].
\bibitem{4'}P. Ho$\check{\textbf{r}}$ava, Phys. Rev. D 79 084008 (2009)
[arXiv:0901.3775 [hep-th]].
\bibitem{5'}P. Ho$\check{\textbf{r}}$ava, JHEP 0903, 020 (2009)
[arXiv:0812.4287 [hep-th]].
\bibitem{6'}P. Ho$\check{\textbf{r}}$ava, Phys.Rev.Lett. 102, 161301 (2009)
[arXiv:0902.3657 [hep-th]].
\bibitem{2}
R. G. Cai, L. M. Cao and N. Ohta, Topological Black Holes in
Ho$\check{\textbf{r}}$ava-Lifshitz Gravity, Phys. Rev. D 80,
024003 (2009) [arXiv:0904.3670 [hep-th]].
\bibitem{3}
R. Banerjee and B. R. Majhi, Connecting anomaly and tunneling
methods for Hawking effect through chirality, Phys. Rev. D 79,
064024 (2009) [arXiv:0812.0497 [hep-th]].
\bibitem{4}
 G. Kunstatter, d-dimensional black hole entropy spectrum from
quasi-normal modes, Phys. Rev. Lett. 90, 161301 (2003)
[arXiv:0212014[gr-qc]].\\
 M. Maggiore, The physical interpretation of
the spectrum of black hole quasinormal modes, Phys. Rev. Lett.
100, 141301 (2008) [arXiv:0711.3145 [gr-qc]].\\ A. J. M. Medved,
D. Martin and M. Visser, Dirty black holes: Quasinormal modes for
squeezed horizons,Class. Quant. Grav. 21, 2393 (2004)
[arXiv:0310097[gr-qc]].
\bibitem{5}
D. Momeni, "Cosmic strings in Ho$\check{\textbf{r}}$ava-Lifshitz
Gravity," to be appear in PLB (2010), [arXiv:0910.594[gr-qc]].
\bibitem{6}
M. R. Setare and D. Momeni , "Plane symmetric solutions in
Ho$\check{\textbf{r}}$ava-Lifshitz theory,"
[arXiv:0911.1877[hep-th]].
\bibitem{7}
D. Blas, O. Pujolas, S. Sibiryakov, JHEP, 10, (2009), 029.
\bibitem{8}
E.Kiritsis, "Spherically symmetric solutions in modified
Ho$\check{\textbf{r}}$ava-Lifshitz gravity,"[arXiv:0911.3164
[gr-qc]].
\bibitem{9'}S. Hod, Phys. Rev. Lett. 81, 4293 (1998).
\bibitem{10'}O. Dreyer, Phys. Rev. Lett. 90, 081301 (2003).
\bibitem{11'}S. Chandrasekhar, The Mathematical Theory of Black Holes, Cambridge University
Press (1983).
\bibitem{12'}K. D. Kokkotas e B. G. Schmidt, Living Reviews in Relativity (1999);
H. -P. Nollert, Class. Quant. Grav. 16, R159-R216 (1999).
\bibitem{9}
M. R. Setare,Area spectrum of extremal Reissner-Nordstroem black
holes from quasi- normal modes, Phys. Rev. D 69, 044016 (2004)
[arXiv:0312061[hep-th]]\\
 M. R. Setare, Near extremal
Schwarzschild-de Sitter black hole area spectrum from quasi-
normal
modes, Gen. Rel. Grav. 37, 1411 (2005) [arXiv:0401063[hep-th]].\\
 M.R. Setare and E. C. Vagenas, Area spectrum of Kerr and extremal
Kerr black holes from quasinormal modes, Mod. Phys. Lett. A 20,
1923 (2005) [arXiv:0401187[hep-th]]\\
 M. R. Setare, Non-rotating BTZ
black hole area spectrum from quasi-normal modes, Class. Quant.
Grav. 21, 1453 (2004) [arXiv:0311221[hep-th]].
\bibitem{10}
A. Kehagias and K. Sfetsos, Phys Lett B 678, 123 (2009)
[arXiv:0905.0477 [hep-th]].
\bibitem{11}
 A. Wang, Y. Wu , JCAP 0907, 012 (2009) [arXiv:0905.4117
[hep-th]].
\bibitem{12}
M. Wang, J. Jing, C. Ding, and S. Chen , "First laws of
thermodynamics in IR Modified Ho$\check{\textbf{r}}$ava-Lifshitz
gravity,"[arXiv:0912.4832 [gr-qc]].
\bibitem{22}R. L. Arnowitt, S. Deser and C. W. Misner, The dynamics of general
relativity, "Gravitation: an introduction to current research",
Louis Witten ed. (Wilew 1962),chapter 7, pp 227-265,
arXiv:gr-qc/0405109.
\bibitem{23}E. Kiritsis and G. Kofinas, Ho$\check{\textbf{r}}$ava-Lifshitz
cosmology,[arXiv:0904.1334 [hep-th].
\bibitem{13}
Y. Sekiwa, Phys.Rev. D 73, 084009 (2006)
\bibitem{15}
J. D. Bekenstein and V. F. Mukhanov, Phys. Lett. B 360, 7 (1995).
$$$$$$$$$$$$$$$$$$$$$$$$$$$$$$$$$$$$$$$$$$$$$$$$$$$$$$$$$$$$






\end{thebibliography}
\end{document}